\documentclass[11pt]{article}

\usepackage{amssymb}

\usepackage{graphicx}

\usepackage[english,francais]{babel}

\newfont{\ensmathquatorze}{msbm10 scaled 1400}
\newfont{\ensmathonze}{msbm10 scaled 1100}
\newfont{\ensmathdix}{msbm10}
\newfont{\ensmathneuf}{msbm10 scaled 833}
\newfont{\ensmathhuit}{msbm10 scaled 694}
\newfam\ensmathfam                        
\textfont\ensmathfam=\ensmathonze        
\scriptfont\ensmathfam=\ensmathdix       
\scriptscriptfont\ensmathfam=\ensmathhuit
\def\ensmf{\fam\ensmathfam\ensmathonze}         

\def\be{\begin{equation}}
\def\ee{\end{equation}}

\def\bea{\begin{eqnarray}}
\def\eea{\end{eqnarray}}
\def\beann{\begin{eqnarray*}}
\def\eeann{\end{eqnarray*}}

\renewcommand{\leq}{\leqslant}
\renewcommand{\geq}{\geqslant}

\newcommand{\ket}[1]{|\kern.3ex#1\kern.3ex\rangle}
\newcommand{\bra}[1]{\langle\kern.3ex #1 \kern.3ex|}
\newcommand{\APPROX}[1]{                
   {{\raisebox{-.3cm}{$\textstyle\simeq$}} \atop {\scriptstyle{#1}}}}

\newcommand{\EXP}[1]{{\mbox{\large e}}^{#1}}         
\renewcommand{\sinh}[1]{\mathop{\mathrm{sh}}\nolimits #1} 

\newcommand{\cotg}{\mathop{\mathrm{cotg}}\nolimits}  
\renewcommand{\min}[2]{\mathop{\mathrm{min}}\nolimits\left( #1 , #2\right)}

\def\NN{{\ensmf N}}                 
\def\RR{{\ensmf R}}                 

\def\I{{\rm i}}                  
\def\D{{\rm d}}                  
\def\Dc{{\rm D}}                 

\newcommand{\drond}[2]{\frac{\partial #1}{\partial #2}} 

\newcommand\ab{{\alpha\beta}}
\newcommand\ba{{\beta\alpha}}
\newcommand\mb{{\mu  \beta}}
\newcommand\bm{{\beta\mu}}
\newcommand\lab{l_{\alpha\beta}}

\begin{document}

\selectlanguage{english}

\title{Scattering theory on graphs}
 
\author{    Christophe Texier$^{(a,b)}$ 
        and Gilles Montambaux$^{(b)}$}

\date{5th July 2001}

\maketitle	

\noindent
{\small
$^{(a)}$Laboratoire de Physique Th\'eorique et Mod\`eles Statistiques.
Universit\'e Paris-Sud, 
B\^at. 100,\\
$^{(b)}$Laboratoire de Physique des Solides. 
Universit\'e Paris-Sud,
B\^at. 510,
F-91405 Orsay Cedex, France.
}

\vspace{0.5cm}

\noindent
\hspace{0.25cm}
E-mail~: 
\begin{minipage}[t]{7cm}
gilles@lps.u-psud.fr\\
texier@ipno.in2p3.fr, texier@lps.u-psud.fr
\end{minipage}

\vspace{0.5cm}

\begin{abstract}
We consider the scattering theory for the Schr\"odinger operator 
$-\Dc_x^2+V(x)$ on graphs made of one-dimensional wires connected to 
external leads. 
We derive two expressions for the scattering matrix on arbitrary graphs.
One involves  matrices that couple arcs (oriented bonds), the other involves 
matrices that couple vertices.
We discuss a simple way to tune the coupling between the graph and the 
leads.
The efficiency of the formalism is demonstrated on a few known examples.
\end{abstract}

\noindent
PACS~: 03.65.Nk, 72.10.Bg, 73.23.-b





\section{Introduction\label{sec:Intro}}

The study of graphs is a vast domain. Spectral theory of the Laplacian on
graphs has been widely studied in the mathematical literature 
\cite{Big76,Chu97,Col98,CveDooSac80}.
Here we are interested on graphs made of one-dimensional wires identified
with finite interval of $\RR$ and being connected at vertices.
A trace formula for the partition function of the Laplace operator on 
such graphs has been derived in a very nice work by J.-P.~Roth
\cite{Rot83,Rot83a} who expressed the partition function in terms of
contributions of periodic orbits.
The study of the Laplace operator on graphs has been shown to be relevant in 
many physical situations. It has been first considered for the study of
organic molecules \cite{RudSch53}.
It has also some interest in the context of superconducting networks
\cite{Ale83}, for the study of adiabatic quantum transport in networks
\cite{AvrRavZur88,AvrSad91} and in the weak localization theory 
\cite{DouRam85,DouRam86,DouRam87c,PasMon99,Pas98}. More precisely 
several physical quantities of weak localization theory are related to the 
spectral determinant of the Laplace operator $S(\gamma)=\det(-\Dc_x^2+\gamma)$,
that can be expressed in terms of the determinant of a $V\times V$ matrix 
$M$ coupling the vertices \cite{PasMon99}.
The relation between $S(\gamma)$ and the trace formula obtained by Roth has 
been examined in \cite{AkkComDesMonTex00}. 
Graphs have also been a subject of several studies in the context of
quantum chaos for their spectral properties 
\cite{KotSmi97,KotSmi99,BerKea99,BerBogKea01} and also their scattering 
properties when they are connected to leads \cite{KotSmi00}.
Scattering theory on graphs has been studied in \cite{KosSch99} and also
frequently used in the context of transport theory for mesoscopic
networks ({\it e.g.} \cite{Sha83,GefImrAzb84,ButImrAzb84,But85})~;
more recently graphs were considered \cite{VidMonDou00} to describe mesoscopic
2-D normal metal networks and superconducting networks realized
experimentally to reveal the so-called Aharonov-Bohm cage effect
\cite{PanAbiSerFouButVid00,NauFaiMaiEti01}.
In order to describe disordered networks, for example to understand how the 
Aharonov-Bohm cage effect is affected by disorder, it is important to have a
simple and efficient formalism which incorporates a potential on the bonds.

In this work we consider the scattering theory for a graph on the bonds of
which lives a potential $V(x)$ and connected to external leads from which
some wave is injected. 
Some spectral properties of the Schr\"odinger operator on graphs have 
already been studied in \cite{Car97}. More recently J.~Desbois
\cite{Des00,Des00a} generalized to the case of the Schr\"odinger operator 
the result for the spectral determinant of the Laplace operator by one of us
and M.~Pascaud \cite{PasMon99}.
Concerning scattering properties, star graphs with potential on the
bonds have been studied in \cite{Exn96}.
The aim of our work is to provide a general and systematic framework to
construct the scattering matrix of a given graph in terms of matrices
encoding the information on the topology and the potential on the graph.

The paper is organized as follows~: in the next section we introduce the basic
definitions.
In section \ref{sec:am} we derive an expression of the scattering matrix of 
the graph in terms of arc matrices (\ref{RES1}).
In section \ref{sec:vm} we take a different point of view and express
the scattering matrix in terms of vertex matrices (\ref{RES2},\ref{RES3}).
Our results generalize the formulae known for the Schr\"odinger operator 
$-\Dc_x^2$ in the absence of scattering by the bonds \cite{AvrSad91,KotSmi99}.
We will see that the second formulation of the scattering matrix with 
vertex matrices offers the advantage of compactness compared to the arc 
matrix formulation.
We discuss, in section \ref{sec:coup}, simple modifications of the  
formalism to introduce tunable couplings between the leads and the graph in
the most efficient way. Simple examples are developed.


\section{Position of the problem \label{sec:def}}

We first define the problem and recall the notations chosen in
\cite{AkkComDesMonTex00,Des00}.
We consider the Schr\"odinger operator 
\be
H=-\Dc_x^2+V(x)
\ee
where $\Dc_x=\D_x-\I A(x)$ is the covariant derivative and the $x$ coordinate
lives on a graph ${\cal G}$ made of $B$ one-dimensional wires connected at
$V$ vertices.
Throughout this paper we will designate the vertices with greek letters
($\alpha$, $\beta$, $\mu$,\ldots). We introduce the $V\times V$-adjacency 
matrix $a_\ab$~: if the vertices $\alpha$ and $\beta$ are linked
by a bond then $a_\ab=1$ and $a_\ab=0$ otherwise. The coordination of vertex
$\alpha$ (number of bonds issuing from the vertex) is 
$m_\alpha=\sum_\beta a_\ab$. 
We call $x_\ab\in[0;\lab]$ the coordinate on the bond $(\ab)$ of length
$\lab$ (note that by definition $x_\ba=\lab-x_\ab$).

The Schr\"odinger operator acts on scalar functions $\psi(x)$ living on  
${\cal G}$ that are represented by a set of $B$ components 
$\psi_{(\ab)}(x_\ab)$ satisfying appropriate boundary conditions at the 
vertices \cite{AvrRavZur88,AvrSad91,Avr95}~:

\noindent ({\it i}) continuity
\be\label{CL1}
\psi_{(\alpha\beta_i)}(x_{\alpha\beta_i}=0)=\psi_\alpha
\ \ \mbox{ for }\ \  i=1,\cdots,m_\alpha
\ee
The indice $\beta_i$ designates a vertex neighbour of vertex $\alpha$~;
the wave function at the vertex is $\psi_\alpha$.

\noindent ({\it ii}) A second condition sufficient to ensure current 
conservation ({\it i.e.} unitarity of the scattering matrix)
\be\label{CL2}
\sum_\beta a_\ab\,\Dc_{x_{\alpha\beta}}\psi_{(\alpha\beta)}(x_\ab=0)
=\lambda_\alpha\psi_\alpha
\:,\ee
where $\lambda_\alpha$ is a real parameter. Due to the presence of the
connectivity matrix $a_\ab$, the sum runs over all neighbouring vertices linked
with vertex $\alpha$. 
To have a better understanding of the physical meaning of the parameter 
$\lambda_\alpha$ we remark that for a vertex of coordination number 2 the 
equation (\ref{CL2}) describes a potential $\lambda_\alpha\delta(x_\ab)$ at 
the position of the vertex $\alpha$. 
Note also that the limit $\lambda_\alpha\to\infty$ corresponds to Dirichlet 
condition $\psi_\alpha=0$ which means that no current is transmitted through 
this vertex.

It is also possible to consider more general boundary conditions than 
(\ref{CL1},\ref{CL2}) and release the continuity condition as it was 
proposed in \cite{KosSch99}.

The magnetic flux along the bond is denoted by
$\theta_\ab=\int_\alpha^\beta\D x\,A(x)=-\theta_\ba$.

We also introduce the notion of {\it arc} which is an {\it oriented bond}. Each
bond $(\ab)$ is associated with two arcs~: $\ab$ and $\ba$. Throughout this
paper we will label the arcs with roman letters ($i$, $j$,\ldots ) and
designate the reversed arc of $i$ with a bar~: $\bar i$.

To describe the potential $V_{(\ab)}(x_\ab)$ on the bond $(\ab)$ it will be
appropriate to introduce reflection and transmission coefficients.
We call $r_\ab(E)$ and $t_\ab(E)$ reflection and transmission probability
amplitudes associated with the transmission from vertex $\alpha$ to
vertex $\beta$ for a plane wave of energy $E$. 
The scattering $2\times2$-matrix for the bond is
\be
\left(
\begin{array}{cc}
r_\ab & t_\ba \\
t_\ab & r_\ba
\end{array}
\right)
\ee

\begin{figure}[!h]
\begin{center}
\includegraphics[scale=0.9]{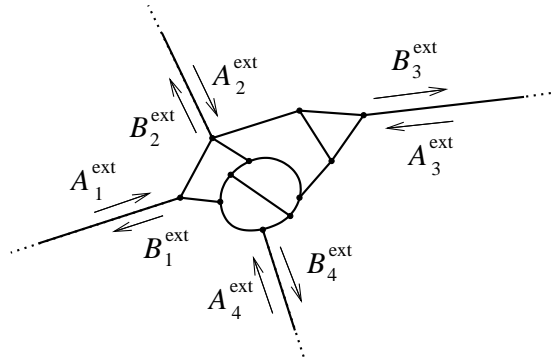}
\end{center}
\caption{A graph made of $B=15$ internal bonds and $V=11$ vertices
connected to $L=4$ external leads.}
\label{fig:gc}
\end{figure}

We are considering a scattering problem, that is we consider a situation
where the graph ${\cal G}$ is connected to $L$ external leads by which
some wave is injected (see figure \ref{fig:gc}). The on-shell scattering
matrix $\Sigma$ is a $L\times L$-matrix that relates the incoming 
amplitudes in the $L$ channels to  the outcoming ones. We call 
$A^{\rm ext}_\alpha$ (resp. $B^{\rm ext}_\alpha$) the incoming (resp. 
outcoming) amplitude on the external lead connected at the vertex $\alpha$. 
By definition~:
\begin{center}
\begin{tabular}{|p{5cm}|}
\hline
\be\label{defscatt}
B^{\rm ext} = \Sigma\: A^{\rm ext}
\:.\ee
\\
\hline
\end{tabular}
\end{center}
The purpose of the paper is to express $\Sigma$ by means of arc
$2B\times2B$-matrices and vertex $V\times V$-matrices. We will generalize 
the expressions known in the absence of potential 
\cite{AvrSad91,KotSmi99,KotSmi00}.

\section{Scattering matrix in terms of arc matrices \label{sec:am}}

In this section we construct the scattering matrix by relating it to 
arc matrices.

\subsection*{Scattering by bonds}

\begin{figure}[!h]
\begin{center}
\includegraphics[scale=0.9]{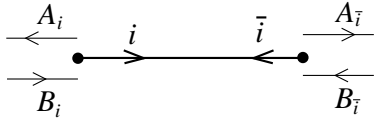}
\end{center}
\caption{the amplitudes on the arcs $i$ and $\bar i$.}
\label{fig:amp}
\end{figure}

We have already explained in section \ref{sec:def} how to describe the
scattering by the potential $V(x)$ on the bonds by $2\times2$-scattering
matrices.
We associate to each internal arc $i$ two amplitudes $A^{\rm int}_i$ and 
$B^{\rm int}_i$ (see figure \ref{fig:amp})~; this means that the component 
$\psi_i(x)$ of the wave function of energy $k^2$ matches with
$A^{\rm int}_i\EXP{-\I kx}+B^{\rm int}_i\EXP{\I kx}$ at the node from which 
arc $i$ issues.
It follows that the amplitudes at the two boundaries of the arc $i$ are
related by~:
\be
\left(
\begin{array}{c}
A^{\rm int}_i \\
A^{\rm int}_{\bar i}
\end{array}
\right)
=\left(
\begin{array}{cc}
r_i & t_{\bar i} \\
t_i & r_{\bar i}
\end{array}
\right)
\left(
\begin{array}{c}
B^{\rm int}_i \\
B^{\rm int}_{\bar i}
\end{array}
\right)
\:,\ee
where $\bar i$ is the reversed arc.
This relation may be more conveniently written in terms of a matrix $R$ that
couples the $2B$ internal arcs~:
\be
A^{\rm int}_i = \sum_j R_{ij} B^{\rm int}_j
\ee
with
\be
R_{ij} = r_i \delta_{i,j} + t_{\bar i} \delta_{\bar i,j}
\ee
where $\delta_{i,j}$ is the Kronecker symbol and indices $i$ and $j$ run
over the labels of the $2B$ internal arcs~:
$i,j\in\{1,\cdots,B,\bar1,\cdots,\bar B\}$.

If there is no potential on the bonds ($V(x)=0$) we recover the $R$-matrix
introduced in \cite{AkkComDesMonTex00}~:
\be
R^0_{ij} = \EXP{\I kl_i} \delta_{\bar i,j}
\:.\ee

The reflection and transmission coefficients characterize the scattering by
the potential alone and if we introduce a magnetic field, the modification
brought is straightforward~: the transmission amplitudes receive additional 
phases
$t_i\to t_i\EXP{\I\theta_i}$ and the reflection amplitudes are not affected
by the magnetic field.
$\theta_{\ab}=\int_\alpha^\beta\D x\, A(x)=-\theta_\ba$ is the magnetic flux 
along arc $\ab$.
The bond scattering matrix then reads~:
\be\label{r1}
R_{ij} = r_i \delta_{i,j} 
       + t_{\bar i}\EXP{-\I\theta_i} \delta_{\bar i,j}
\ee
This matrix can also be written in a vertex notation (we identify $i$ with 
$\ab$ and $j$ with $\mu\nu$)~:
\be\label{r2}
R_{\alpha\beta,\mu\nu} = 
a_{\alpha\beta}a_{\mu\nu}
\left( r_\ab                   \delta_{\alpha\mu}\delta_{\beta\nu}
     + t_\ba\EXP{\I\theta_\ba} \delta_{\alpha\nu}\delta_{\beta\mu} 
\right)
\:,\ee
where the adjacency matrix elements $a_\ab$ and $a_{\mu\nu}$ ensure that
$\alpha$ and $\beta$ are connected by a bond, as well as $\mu$ and $\nu$.

\subsection*{Scattering by vertices}

The bond scattering matrix only couples amplitudes $A^{\rm int}$ and
$B^{\rm int}$ associated with internal arcs. 
On the other hand some vertices ($L$) couple internal bonds and external 
leads. We write the wave function on the lead connected to the vertex 
$\alpha$ as (see figure \ref{fig:gc})~: 
\be
\psi_{{\rm lead}\:\alpha}(x) = A^{\rm ext}_{\alpha}\EXP{-\I k x}
+B^{\rm ext}_{\alpha}\EXP{\I k x}
\ee
($x=0$ coincides with the vertex).
Since we have to introduce only one pair of amplitudes $A^{\rm ext}_{\alpha}$,
$B^{\rm ext}_{\alpha}$ per external lead, this means that each lead is 
described by one arc only. Adopting this convention implies that we are now 
dealing with $2B+L$ arcs.
We group the internal and external amplitudes in a unique vector~:
\be
A = 
\left(
\begin{array}{c}
A^{\rm int} \\ \hline
A^{\rm ext}
\end{array}
\right)
\hspace{0.5cm}\mbox{ and }\hspace{0.5cm}
B = 
\left(
\begin{array}{c}
B^{\rm int} \\ \hline
B^{\rm ext}
\end{array}
\right)
\:.\ee

If we consider a given vertex $\alpha$ of coordination $m_\alpha$, it
follows from (\ref{CL1},\ref{CL2}) that the $m_\alpha$ incoming amplitudes 
$A_i$ at the vertex are related to the outgoing  amplitudes $B_i$ by a
$m_\alpha\times m_\alpha$ unitary matrix $Q_\alpha$
whose diagonal elements are $\frac{2}{m_\alpha+\I\lambda_\alpha/k}-1$, all
other being $\frac{2}{m_\alpha+\I\lambda_\alpha/k}$.
We call $Q$ the  $(2B+L)\times(2B+L)$-vertex scattering matrix of the whole 
graph with leads \cite{AkkComDesMonTex00}~:
\be
B_i = \sum_j Q_{ij} A_j
\ee
with~:
\bea
Q_{ij} & = & \frac{2}{m_\alpha+\I\lambda_\alpha/k}-1
             \hspace{0.5cm} \mbox{ if } i=j
	     \mbox{ ($i$ issues from the vertex $\alpha$)}  \label{q1}\\
       & = & \frac{2}{m_\alpha+\I\lambda_\alpha/k}
             \hspace{1.15cm} \mbox{ if } i\neq j
	     \ \mbox{ both issuing from the vertex $\alpha$} \label{q2}\\
       & = & 0 \hspace{3cm}\mbox{ otherwise} \label{q3}
\:.\eea
We can also write the matrix elements for the internal arcs in a vertex 
notation~:
\be\label{q4}
Q_{\alpha\beta,\mu\nu}=a_{\alpha\beta}a_{\mu\nu}
\delta_{\alpha\mu}\left(\frac{2}{m_\alpha+\I\lambda_\alpha/k}
-\delta_{\beta\nu}\right)
\:.\ee
All the information on the topology of the graph is encoded in the matrix $Q$.

\subsection*{Scattering by the full graph}

We have seen that the scattering by bonds relates internal amplitudes~:
\be\label{bs}
A^{\rm int} = R\:B^{\rm int}
\ee
and the scattering by vertices all amplitudes~:
\be\label{vs}
B = Q\: A
\:.\ee
We separate the $Q$ matrix into 4 block matrices~:
\be
Q = 
\left(
\begin{array}{c|c}
Q^{\rm int} & \tilde Q^{\rm T} \\ \hline 
\tilde Q   & Q^{\rm ext}
\end{array}
\right)
\ee
where $Q^{\rm T}$ is the transposed matrix ($Q^{\rm int}$ is a
$2B\times2B$-matrix, $Q^{\rm ext}$ is a $L\times L$-matrix and
$\tilde Q$ is a $L\times2B$-matrix). 
In the following we will always choose to write the matrix $Q$ according
to this structure.

Equation (\ref{vs}) becomes~:
\bea\label{vs1}
B^{\rm int} &=& Q^{\rm int}\, A^{\rm int} + \tilde Q^{\rm T}\, A^{\rm ext} \\
\label{vs2}
B^{\rm ext}&=& \tilde Q\, A^{\rm int}   + Q^{\rm ext}\,A^{\rm ext}
\:.\eea
We can now eliminate in (\ref{bs},\ref{vs1},\ref{vs2}) the internal
amplitudes and relate $A^{\rm ext}$ to $B^{\rm ext}$. Therefore we obtain the
scattering matrix of the graph~:
\begin{center}
\begin{tabular}{|p{8cm}|}
\hline
\be\label{RES1}
\Sigma = Q^{\rm ext} + 
\tilde Q \, ({R^{\dagger} - Q^{\rm int}})^{-1} \, \tilde Q^{\rm T}
\:.\ee
\\
\hline
\end{tabular}
\end{center}
We have used the unitarity of $R$~: 
\be
(R^{-1})_{ij} =(R^{\dagger})_{ij} = r_i^* \delta_{i,j} 
              + t_{i}^* \EXP{-\I\theta_i} \delta_{\bar i,j}
\:.\ee
The expression (\ref{RES1}) generalizes the result known in the absence of
potential \cite{KotSmi00}.

\subsection*{Example}

\begin{figure}[!h]
\begin{center}
\includegraphics[scale=0.9]{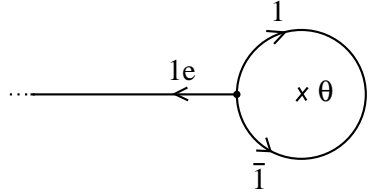}
\end{center}
\caption{Scattering in a ring pierced by a magnetic flux.}
\label{fig:ring}
\end{figure}

As an example we consider the scattering on the ring of perimeter $l$
pierced by a flux $\theta$ (figure \ref{fig:ring}) without potential.
This graph possesses one internal bond (arcs $1$ and $\bar1$)~; the external
lead is associated with an arc called $1e$. 
The bond scattering matrix (\ref{r1},\ref{r2}) is~:
\be
R = 
\left(
\begin{array}{cc}
 0                     & \EXP{\I k l -\I\theta} \\ 
\EXP{\I k l +\I\theta} & 0 
\end{array}
\right)
\ee
and the vertex scattering matrix (\ref{q1},\ref{q2},\ref{q3},\ref{q4}), 
expressed in a basis of arcs $\{1,\bar1,1e\}$ (see figure \ref{fig:ring}) 
for $\lambda_\alpha=0$, reads
\be\label{MQ3}
Q=
\left(
\begin{array}{cc|c}
-1/3 &  2/3 &  2/3  \\ 
 2/3 & -1/3 &  2/3  \\ \hline
 2/3 &  2/3 & -1/3
\end{array}
\right)
\:.\ee

\noindent
Applying (\ref{RES1}) we find
\be\label{sring}
\Sigma = - \frac{3\EXP{\I k l} -4\cos\theta + \EXP{-\I k l}}
                {3\EXP{-\I k l}-4\cos\theta + \EXP{\I k l}}
\:.\ee
With one lead, the scattering matrix is given by a unique phase~:
$\Sigma(E) = \EXP{\I\delta(E)}$ with
\be\label{sring2}
\cotg\frac{\delta}{2} = \frac{\sin kl}{2(\cos\theta-\cos kl)}
\:;\ee
we recover a result obtained by relating in the one channel case the 
scattering matrix to ratio of spectral determinants 
\cite{AkkComDesMonTex00}.

\subsection*{Remark~: multichannel wires}

We remark that the formulation of the scattering in terms of arc matrices
can be generalized for multichannel wires~: the matrix elements 
$Q_{ij}$ and $R_{ij}$ would then become submatrices coupling channels.

\subsection*{Multiple scattering expansion}

It is sometimes interesting to expand the quantities of interest in terms of
contributions of paths in the graph (we call {\it path} an ordered set of
arcs). 
Since the matrices $Q$ and $R$ contain the scattering amplitudes on vertices
and bonds, respectively, it is obvious that the expansion of (\ref{RES1}) 
expresses the contributions of paths to the transmission amplitudes from one 
lead to another~: 
\be
\Sigma = Q^{\rm ext} 
+ \tilde Q\, R\,                   \tilde Q^{\rm T}
+ \tilde Q\, R Q^{\rm int} R\,     \tilde Q^{\rm T} + \cdots 
+ \tilde Q\, R (Q^{\rm int} R)^n\, \tilde Q^{\rm T} + \cdots
\ee
The first term is associated with transmission from leads without entering
the graph. The term $\tilde Q R\tilde Q^{\rm T}$ corresponds to paths that
contain only one bond of the graph. More generally, the element
$(\tilde Q R(Q^{\rm int}R)^n\tilde Q^{\rm T})_{ij}$ is the sum of all 
amplitudes associated to the paths going from lead $j$ to lead $i$, and made 
of $n+1$ internal arcs.


\section{Scattering matrix in terms of vertex matrices \label{sec:vm}}

The approach presented in the previous section has the advantage to consider
only scattering matrices for bonds and vertices but presents the disagreement
to manipulate rather big matrices ($2B\times2B$).
In this section we follow a different methodology by constructing the
stationary scattering states in the graph which leads to deal with vertex
matrices ($V\times V$) usually smaller.

For convenience we label the vertices connected to leads with the $L$ first
indices~: $\alpha=1,\cdots,L$, however the final result will be completely 
independent of the way the basis of vertices is organized.

We introduce the $L\times V$-matrix $W$ \cite{KotSmi99} containing the
information about the way the graph is connected~:
$W_\ab=\delta_\ab$ with $\alpha\in\{1,\cdots,L\}$ and
$\beta\in\{1,\cdots,V\}$~:
\be
W = 
\left(
\begin{array}{cccc|ccc}
1      & 0      & \cdots & 0      & 0      & \cdots & 0 \\
0      & 1      & \cdots & 0      & 0      & \cdots & 0 \\
\vdots & \vdots & \ddots & \vdots & \vdots & \ddots & \vdots \\
0      & \cdots & 0      & 1      & 0      & \cdots & 0 \\
\end{array}
\right)
\ee

We now turn to the construction of the stationary scattering states
$\psi^{(\alpha)}(x)$ of energy $k^2$ which describes a plane wave entering 
the graph from the lead connected at vertex $\alpha$ and being scattered by 
the graph into all leads.
We consider the case without magnetic field since the addition of a magnetic
field is straightforward by adding the appropriate phases in the
transmission coefficients of the bonds.

On the lead connected to vertex $\mu$, the wave function is~:
\be\label{wfl}
\psi^{(\alpha)}_{{\rm lead}\:\mu}(x) = 
\delta_{\mu\alpha} \EXP{-\I kx} + \Sigma_{\mu\alpha} \EXP{\I kx}
\:,\ee
with $x\in[0;+\infty[$.

To construct the wave function on the internal bond $(\mb)$ of the graph, 
it is convenient to introduce the two linearly independent solutions
$f_\mb(x_\mb)$ and $f_\bm(x_\mb)$ of the differential equation
\be\label{Schroed}
\left( -\D_{x_\mb}^2 + V_{(\mb)}(x_\mb) + \gamma \right) f(x_\mb) = 0
\:\ee
for $x\in[0,l_\mb]$~,
satisfying the following boundary conditions at the edges of the
interval~:
\be
\left\{ 
\begin{array}{l}
f_\mb(\mu)   = 1 \\ [0.1cm]
f_\mb(\beta) = 0
\end{array}\right.
\hspace{0.5cm} \mbox{and} \hspace{0.5cm}
\left\{ 
\begin{array}{l}
f_\bm(\mu)   = 0 \\ [0.1cm]
f_\bm(\beta) = 1
\end{array}\right.
\ee
We follow here the construction of the spectral determinant for the
Schr\"odinger operator in \cite{Des00}. To lighten the expressions
we have introduced the obvious notation $f(\mu)\equiv f(x_\mb=0)$ and 
$f(\beta)\equiv f(x_\mb=l_\mb)$.
The spectral parameter is~:
\be
\gamma = - k^2 -\I 0^+
\:.\ee

For example, if $V_{(\mb)}(x)=0$ the two functions are~: 
$f_\mb(x_\mb)=\frac{\sinh\sqrt{\gamma}(l_\mb-x_\mb)}{\sinh\sqrt{\gamma}l_\mb}$
and 
$f_\bm(x_\mb)=\frac{\sinh\sqrt{\gamma}x_\mb}{\sinh\sqrt{\gamma}l_\mb}$.

We call $\psi^{(\alpha)}_\mu$ the wave function at the vertex $\mu$ when the
plane wave is injected at vertex $\alpha$. 
The solution of the Schr\"odinger equation (\ref{Schroed}) on the bond $(\mb)$
\be\label{wfb}
\psi^{(\alpha)}_{(\mb)}(x_\mb) = 
\psi^{(\alpha)}_\mu\,f_\mb(x_\mb) + \psi^{(\alpha)}_\beta\,f_\bm(x_\mb)
\ee
already satisfies the continuity condition (\ref{CL1}).

If we impose the condition (\ref{CL1}) for the wave function on the lead 
($\psi^{(\alpha)}_{{\rm lead}\:\mu}(0)=\psi^{(\alpha)}_\mu$) we get~:
\be\label{eq1}
\delta_{\mu\alpha}  + \Sigma_{\mu\alpha} = \psi^{(\alpha)}_\mu
\hspace{0.5cm} \mbox{ for } \mu=1,\cdots,L
\:.\ee
The solution $\psi^{(\alpha)}(x)$ must also satisfy the condition
(\ref{CL2}), that is~:
\be
\sum_\beta a_\mb \frac{\D \psi^{(\alpha)}_{(\mb)} }{\D x_\mb}(\mu)
+ (W^{\rm T}W)_{\mu\mu}\, 
\frac{\D \psi^{(\alpha)}_{{\rm lead}\:\mu}}{\D x}(\mu)
=\lambda_\mu\psi^{(\alpha)}_\mu
\hspace{0.5cm} \mbox{ for } \mu=1,\cdots,V
\:.\ee
The $(W^{\rm T}W)_{\mu\mu}$ ensures that this contribution to current from
leads vanishes if $\mu$ is an internal vertex. 
This equation can be rewritten as
\be\label{eq2}
(W^{\rm T}W)_{\mu\mu}\,\left(\delta_{\mu\alpha}  - \Sigma_{\mu\alpha}\right)
= \sum_\beta M_{\mb} \psi^{(\alpha)}_\beta
\hspace{0.5cm} \mbox{ for } \mu=1,\cdots,V
\:,\ee
where $M$ is the matrix appearing in the expression of the spectral 
determinant\footnote{
We have used the fact that the Wronskian is equal to~:
$$
{\cal W}[f_\mb,f_\bm] = f_\mb \frac{\D f_\bm}{\D x_\mb}
      - \frac{\D f_\mb}{\D x_\mb}f_\bm
      = \frac{\D f_\bm}{\D x_\mb}(\mu) = - \frac{\D f_\mb}{\D x_\mb}(\beta)
\:.$$
}\cite{Des00}
\be\label{MJean}
M_\mb(\gamma) = \frac1{\sqrt{\gamma}}
\left(
  \delta_\mb
  \left[
    \lambda_\mu - 
    \sum_\nu a_{\mu\nu} \frac{\D f_{\mu\nu}}{\D x_{\mu\nu}}(\mu)
  \right]
  + a_\mb \frac{\D f_\mb}{\D x_\mb}(\beta)
\right)
\:.\ee

If we consider $\psi^{(\alpha)}_\mu$ as the matrix elements $(\mu,\alpha)$
of a $V\times L$ matrix $\Psi$, (\ref{eq1},\ref{eq2}) can be rewritten in a 
matrix form~:
\bea\label{eq3}
1 + \Sigma             &=& W \Psi \\
\label{eq4}
W^{\rm T} (1 - \Sigma) &=& M \Psi
\:.\eea
We obtain the scattering matrix by eliminating $\Psi$ in 
(\ref{eq3},\ref{eq4}) (with the help of the identity recalled in appendix
\ref{InvMat}). Finally we get~:
%
\begin{center}
\begin{tabular}{|p{8cm}|}
\hline
\be\label{RES2}
\Sigma = -1 + 2 \, W \left( M + W^{\rm T}W\right)^{-1} W^{\rm T}
\:.\ee
\\
\hline
\end{tabular}
\end{center}

The last step is to relate the matrix $M$ to the reflection and transmission
coefficients of the bonds. For this purpose we note that we could have
chosen a different basis of solutions of equation (\ref{Schroed}) to
construct the stationary state (\ref{wfb}) on the bond.
In particular we could have chosen the right $\phi_\bm(x)$ and left
$\phi_\mb(x)$ stationary scattering states associated with the potential 
$V_{(\mb)}(x)$ of the bond solely. If we think at the bond potential
$V_\bm(x)$ with support $[0,l_\mb]$ embedded in an infinite line ($\RR$) 
these states would be written out of the interval as~:
\be
\begin{array}{lcll}
\phi_\mb(x) & = & \EXP{\I kx} + r_\mb\,\EXP{-\I kx} & \mbox{ for } x\leq0 
                                                     \vspace{0.15cm} \\
            & = & t_\mb\,\EXP{\I k(x-l_\mb)}        & \mbox{ for } x\geq l_\mb
                                                     \vspace{0.15cm} \\
\phi_\bm(x) & = & t_\bm\,\EXP{-\I kx}               & \mbox{ for } x\leq0 
                                                     \vspace{0.15cm}  \\
            & = & \EXP{-\I k(x-l_\mb)} + r_\bm\,\EXP{\I k(x-l_\mb)} 
	                                           & \mbox{ for } x\geq l_\mb
\end{array}
\ee
It is easy to see that the functions $f_\mb(x)$ are related to those
stationary scattering states by~:
\be
f_\mb(x_\mb) = \frac{(1+r_\bm)\,\phi_\mb(x_\mb) - t_\mb\,\phi_\bm(x_\mb)}
                    {(1+r_\mb)(1+r_\bm)-t_\mb \, t_\bm}
\:.\ee
Then
\be\label{frt1}
\frac{\D f_\mb}{\D x_\mb}(\mu)
= \I k\frac{(1-r_\mb)(1+r_\bm)+t_\mb\,t_\bm}{(1+r_\mb)(1+r_\bm)-t_\mb\,t_\bm}
\ee
and
\be\label{frt2}
\frac{\D f_\mb}{\D x_\mb}(\beta)
= \I k\frac{2\,t_\mb}{(1+r_\mb)(1+r_\bm)-t_\mb\,t_\bm}
\:.\ee
We can now express the matrix $M$ for $\gamma=-k^2-\I 0^+$ in terms of bond 
reflections and transmissions~:
\begin{center}
\begin{tabular}{|p{13cm}|}
\hline
\bea\label{RES3}
M_\ab &=&
\delta_\ab\left(\I\frac{\lambda_\alpha}{k}
+ \sum_\mu a_{\alpha\mu}
\frac{(1-r_{\alpha\mu})(1+r_{\mu\alpha})+t_{\alpha\mu}\,t_{\mu\alpha}}
     {(1+r_{\alpha\mu})(1+r_{\mu\alpha})-t_{\alpha\mu}\,t_{\mu\alpha}}
\right)
\nonumber \\ && \hspace{1cm}
- a_\ab\frac{2\,t_\ab}{(1+r_\ab)(1+r_\ba)-t_\ab\,t_\ba}
\:.\eea
\\
\hline
\end{tabular}
\end{center}
This equation with (\ref{RES2}) generalizes the result known in the 
absence of the potential \cite{AvrSad91,KotSmi99}.
In the appendix \ref{appA} we rewrite the matrix $M$ with real parameters 
replacing the complex reflection and transmission coefficients of the bonds,
and in the appendix \ref{sec:loop} we discuss how it is modified if the 
graph contains loops that we don't want to describe with several vertices.

We repeat that the addition of a magnetic field implies the substitution~:
$t_\ab\to t_\ab\EXP{\I\theta_\ab}$, the reflections being unchanged.

Note that if $V(x)=0$ we have $r_\ab=0$ and 
$t_\ab=\EXP{\I kl_\ab+\I\theta_\ab}$ and we recover the well-known matrix 
\cite{RudSch53,Ale83,Avr95} that appears in the search of the eigenvalues 
of the closed graph (if $\lambda_\alpha=0$)~:
\be
M^0_\ab = \I\,\delta_\ab\sum_\mu a_{\alpha\mu} \cotg kl_{\alpha\mu}
-a_\ab\frac{\I\,\EXP{\I\theta_\ab}}{\sin kl_\ab}
\:.\ee

\subsection*{Example}

We consider again the case of the ring (figure \ref{fig:ring})~;
this example has been studied in \cite{AvrSad91}. The graph
can be described with only one vertex to which one loop is attached. In this 
case the matrix $M$ reduces to a scalar (see \cite{Pas98,AkkComDesMonTex00}
and appendix \ref{sec:loop})~:
\be\label{Mring}
M = 2\I \left(\cotg kl - \frac{\cos\theta}{\sin kl}\right)
\:;\ee
the matrix $W$ reduces to $1$ and we recover straightforwardly from 
(\ref{RES2}) the result (\ref{sring})~:
\be
\Sigma = \frac{\I\sin kl+2(\cos kl-\cos\theta)}
              {\I\sin kl-2(\cos kl-\cos\theta)}
\:.\ee

\subsection*{Remark~: spectral determinant}

Note that the spectral determinant $S(\gamma)=\prod_n(E_n+\gamma)$ characterizing the spectrum of the isolated graph can also be expressed in terms
of the reflection and transmission coefficients by using equations
(\ref{frt1},\ref{frt2},\ref{RES3}) with the result of J.~Desbois 
\cite{Des00}~:
\be
S(\gamma)=\gamma^{V/2}\prod_{(\ab)}
\left(\frac{\D f_\ba}{\D x_\ab}(\alpha)\right)^{-1}\det M(\gamma)
\:.\ee


\section{Tuning the coupling of the graph to the leads \label{sec:coup}} 

In this section we consider the situation where a graph ${\cal G}$  can be 
decoupled from the leads at which it is connected by tuning some parameters.
A way to proceed is to add a bond with a tunable transmission between each 
lead and the corresponding vertex to which it is plugged in (figure 
\ref{fig:gac})~; this can be described with the formalism we have presented 
above in the two previous sections but requires to consider a new 
graph $\tilde{\cal G}$ with $V+L$ vertices and $B+L$ bonds 
(if ${\cal G}$ has $V$ vertices, $B$ bonds and $L$ leads). 
The purpose of this section is to demonstrate that the problem
can be reduced, in the sense that we can keep considering the original graph
${\cal G}$ with $V$ vertices and $B$ bonds, provided some modifications of
the above formalism are made~:
({\it i}) in the ``arc matrices'' formulation we have to modify the vertex
scattering matrix for vertices connected to leads.
({\it ii}) In the ``vertex matrices'' formulation, formulae
(\ref{RES2},\ref{RES3}) still hold using the matrix $M$ of ${\cal G}$ if we 
modify the matrix $W$ in a way that appears to be very natural.

\subsection*{A vertex scattering matrix including arbitrary coupling of one 
arc}

\begin{figure}[!h]
\begin{center}
\includegraphics[scale=0.8]{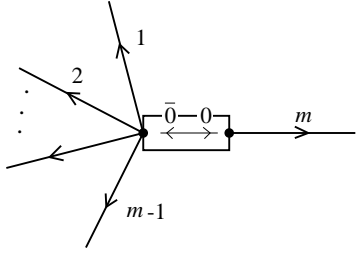}
\end{center}
\caption{The box on the arc $0$ represents a potential characterized by the 
scattering matrix (\ref{smotb}). The arc $m$ can be decoupled
from the other external arcs by tuning the transmission through the bond 
$(0)$.}
\label{fig:newver}
\end{figure}

We construct the scattering matrix of the graph of figure \ref{fig:newver}
made of one bond (two arcs $0$ and $\bar0$). To describe the scattering on the 
bond $(0)$ we choose a simple bond scattering matrix (\ref{r1})
\be\label{smotb}
R = \left(\begin{array}{cc}
\cos\xi & \sin\xi \\  \sin\xi & -\cos\xi
\end{array}\right)
\ee
that allows to tune the transmission probability through the bond~:
$T=\sin^2\xi$. At one side of the bond, 
$m-1$ arcs are connected and only one at the other side. The scattering matrix
we will obtain is the scattering matrix for a vertex with $m$ arcs
among which one can be disconnected by tuning the parameter $\xi$, all other 
arcs being equivalent.

In the basis of arcs $\{0,\bar0,1,2,\cdots,m\}$, the matrix $Q$
(\ref{q1},\ref{q2},\ref{q3}) is~: 
\be
Q=
\left(\begin{array}{cc|ccccc}
\frac{2}{m} -1 
  & 0 
  & \frac{2}{m} 
  & \frac{2}{m} 
  & \cdots 
  & \frac{2}{m}   
  & 0 \\
0          
  & 0      
  & 0         
  & 0         
  & \cdots 
  & 0         
  & 1 \\ \hline
\frac{2}{m}    
  & 0      
  & \frac{2}{m} -1
  & \frac{2}{m}   
  & \cdots & \frac{2}{m}   
  & 0 \\ [0.1cm]
\frac{2}{m}    
  & 0      
  & \frac{2}{m}   
  & \frac{2}{m} -1
  & \cdots 
  & \frac{2}{m}   
  & 0 \\
\vdots     
  & \vdots 
  & \vdots    
  & \vdots    
  &\ddots  
  & \vdots    
  & \vdots\\
\frac{2}{m}    
  & 0      
  & \frac{2}{m}   
  & \frac{2}{m}   
  & \cdots 
  & \frac{2}{m} -1
  & 0 \\
0          
  & 1      
  & 0         
  & 0         
  & \cdots       
  & 0         
  & 0
\end{array}\right)
\:.\ee
The matrix has been written with the parameter $\lambda=0$ to lighten the 
expressions. This parameter is straightforwardly re-introduced by performing
the following substitution~:
\be
m \to m+\I\frac{\lambda}{k}
\:.\ee
We now use the equation (\ref{RES1}) to express the $m\times m$ scattering 
matrix of the
graph~:
\be\label{qt1}
\Sigma = 
\left(
\begin{array}{c|c}
 \rho & \tau^{\rm T} \\ \hline
 \tau & \rho' 
\end{array}
\right)
\:,\ee
where $\rho$ is a $(m-1)\times(m-1)$ matrix, $\tau$ a line vector of 
dimension $m-1$ and $\rho'$ a number~:
\bea
\rho_{ij} &=& \frac{2}{m}\,
 \frac{1+\cos\xi}{1+\left(1-\frac{2}{m}\right)\cos\xi}
              - \delta_{i,j}\:, \\
\tau_i    &=&   \frac{2}{m}\, 
   \frac{\sin\xi}{1+\left(1-\frac{2}{m}\right)\cos\xi}\:,\\
\rho'     &=& - \frac{ 1 -\frac{2}{m}+\cos\xi}
                     {1+\left(1-\frac{2}{m}\right) \cos\xi}
\:;\eea
the indices $i,j$ run over the first $m-1$ equivalent arcs.

A more convenient parametrization is obtained by relating $\xi\in]-\pi,\pi]$
to a parameter $w\in\RR$~:
\be\label{relox}
w = \tan(\xi/2)
\:.\ee
We emphasize that the parameter $w$ characterizes only the scattering
through the bond $(0)$ (figure \ref{fig:newver}).
With this new parameter the scattering matrix takes the simpler form~:
\bea
\label{qt2}
\rho_{ij} &=& \frac{2}{m^\star} - \delta_{i,j}\:, \\ [0.1cm]
\label{qt3}
\tau_i    &=& \frac{2\,w}{m^\star}\:, \\ [0.1cm]
\label{qt4}
\rho'     &=& \frac{2\,w^2}{m^\star} -1
\:,\eea
where we have re-introduced the parameter $\lambda$ in
\be
m^\star = m-1+w^2+\I\frac{\lambda}{k}
\:.\ee
$m^\star$ plays the role of an effective coordination number. The expressions 
(\ref{qt2},\ref{qt3},\ref{qt4}) generalize the vertex scattering matrix 
introduced in \cite{Rot83} to the case of tunable couplings to the leads.
These transmission coefficients were used to calculate the weigths of the 
periodic orbits involved in the trace formula \cite{Rot83,Rot83a} and later 
in \cite{KotSmi97,KotSmi99,KotSmi00}. 

Let us examine several limiting cases to have a better understanding of the
role of the parameter $w$~:

\noindent$\bullet$
If $w=1$, the matrix $\Sigma$ is the symmetric $m\times m$ scattering 
matrix for a vertex of coordinence $m$ given by 
(\ref{q1},\ref{q2},\ref{q3},\ref{q4}).
In this case the transmission of the bond $(0)$ is $T=1$.

\noindent$\bullet$
If $w=0$, the last arc is decoupled from the others and no current
is transmitted to this arc. The scattering between the $m-1$ other arcs, 
described by the $(m-1)\times(m-1)$ matrix $\rho$,
is given by the usual scattering matrix (\ref{q1},\ref{q2},\ref{q3},\ref{q4})
for a coordinence $m-1$.

\noindent$\bullet$
If $w=\sqrt{m-1}$ and $\lambda=0$, the scattering matrix coincides with the 
one introduced by Shapiro \cite{Sha83} up to an inessential change of the 
sign of $\rho$  (this case corresponds to $\cos\xi=-1+\frac2m$, {\it i.e.} a 
transmission $T=\frac{4(m-1)}{m^2}$).

\noindent$\bullet$
If $w=\pm\infty$, all the arcs are decoupled~: $\rho_{ij}=-\delta_{i,j}$,
$\tau_i=0$ and $\rho'=1$.
From the point of view of the $m-1$ first arcs, this limit is equivalent 
to $\lambda=\pm\infty$.

\vspace{0.25cm}

Here we have given a generalization of the scattering matrix proposed in
\cite{ButImrAzb84} for the case of coordination $m=3$ and $\lambda=0$. 
A generalization to any $m$ of the parametrization of B\"uttiker 
{\it et al.} is~:
\be\label{But}
\Sigma=
\left(
\begin{array}{ccc|c}
 b               & b               & \cdots & \sqrt{\epsilon} \\
 b               & b               & \cdots & \sqrt{\epsilon} \\ 
 \vdots          & \vdots          & \ddots & \vdots          \\ \hline
 \sqrt{\epsilon} & \sqrt{\epsilon} & \cdots & c
\end{array}
\right) -1
\:,\ee
where $b=\frac{1}{m-1}\left(1+\sqrt{1-(m-1)\epsilon}\right)$
and $c=2-(m-1)b=1-\sqrt{1-(m-1)\epsilon}$. The relation with our 
parametrization with $w$ is given by~:
$\sqrt{\epsilon}=\frac{2w}{m^\star}$ (then $b=\frac{2}{m^\star}$), valid
for $\lambda=0$.
Note however that the parametrization with $\epsilon\in[0,1/(m-1)]$ does not 
allow to cover the full range of the parameter $w\in\RR$, but only the 
interval $w\in[0,\sqrt{m-1}]$.

\subsection*{Scattering matrix of the graph with arbitrary coupling to the
leads}

\begin{figure}[!h]
\begin{center}
\includegraphics[scale=0.8]{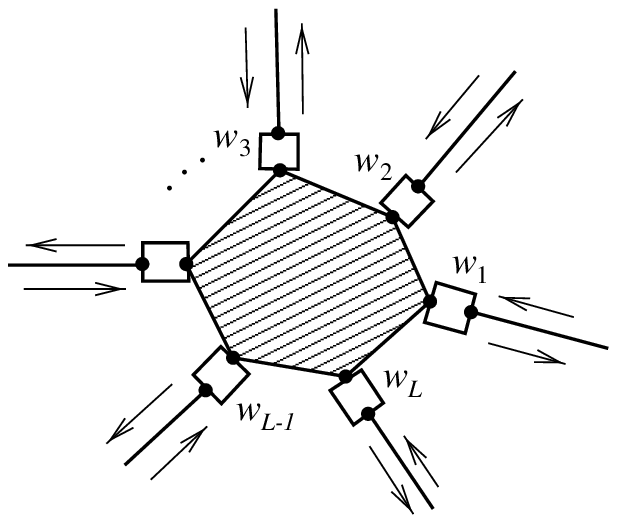}
\end{center}
\caption{Graph with arbitrary coupling to leads. 
The couplings $w_\alpha$'s are represented by boxes~; we recall that they are 
defined by $w_\alpha=\tan(\xi_\alpha/2)$ where the transmission through the 
box is $T_\alpha=\sin^2\xi_\alpha$. The dashed area schematizes the internal
structure of the graph.}
\label{fig:gac}
\end{figure}

We now consider the graph ${\cal G}$ of figure \ref{fig:gac}.
Each external lead is connected to vertices $\alpha\in\{1,2,\cdots,L\}$ of
the graph through a barrier which is described by a parameter 
$w_\alpha\in\RR$~; we call those vertices ``external vertices''.
The scattering matrix of the full graph can be constructed with (\ref{RES1}).
Let us discuss the structure of the vertex scattering matrix. 
$Q$ couples arcs issuing from the same vertex~; to help the discussion, let 
us imagine for a moment that the basis of arcs is organized so that the 
arcs issuing from the same vertex are grouped. The matrix $Q$ is a 
block diagonal matrix in such a basis. As above we call $Q_\alpha$ the 
$m_\alpha\times m_\alpha$ block coupling the arcs issuing from the vertex 
$\alpha$. The blocks related to internal vertices 
$\alpha\in\{L+1,\cdots,V\}$ are unchanged, still given by 
(\ref{q1},\ref{q2},\ref{q3}),
whereas the blocks coupling arcs issuing from external vertices 
$\alpha\in\{1,\cdots,L\}$ are now given by 
(\ref{qt1},\ref{qt2},\ref{qt3},\ref{qt4})~:
\be\label{vce}
Q_\alpha= \frac{2}{m_\alpha^\star}
\left(\begin{array}{cccc|c}
1        & 1        & \cdots & 1        & w_\alpha \\
1        & 1        & \cdots & 1        & w_\alpha \\
\vdots   & \vdots   & \ddots & \vdots   & \vdots  \\
1        & 1        & \cdots & 1        & w_\alpha \\ \hline
w_\alpha & w_\alpha & \cdots & w_\alpha & w_\alpha^2  
\end{array}\right)
-1
\:,\ee
where $m_\alpha^\star\equiv m_\alpha-1+w_\alpha^2+\I\lambda_\alpha/k$.
The introduction of the couplings in this way does not increase the size 
of the matrices we have to deal with by using (\ref{RES1}).

We would like now to generalize formula (\ref{RES2}) without increasing the
difficulty of the calculation of $\Sigma$. 
The construction of the scattering matrix $\Sigma$ using vertex matrices has 
used as a basic ingredient the continuity of the wave function at the 
vertices. If we now describe the scattering at the external vertices with 
(\ref{vce}), this means that the wave function is not anymore continous at 
those vertices due to their internal structure (but still continuous at 
vertices inside the graph).
For a moment we focus on the vertex $\alpha$ with $m_\alpha$ arcs 
among which $m_\alpha-1$ are internal arcs of the graph, the remaining arc 
being a lead.
We call $A_1,A_2,\cdots,A_{m_\alpha-1}$ the $m_\alpha-1$ incoming 
amplitudes from the graph and $A_{m_\alpha}$ the incoming amplitude from the 
lead. 
Let us examine the value of the wave function on the arcs $i$~: 
$\psi_i(x)=A_i\EXP{-\I kx}+B_i\EXP{\I kx}$. 
We have $B_i = \sum_{j=1}^{m_\alpha} (Q_\alpha)_{ij} A_j$~; 
on the arc $i$, if $x\to0$ the wave function goes to 
$\psi_i(0)=A_i+B_i$. It follows from the expression (\ref{vce}) that we
still have the continuity for the wave function on the arcs inside the graph
\be
\psi_1(0)=\cdots=\psi_{m_\alpha-1}(0) =
\frac{2}{m_\alpha^\star}\:{\cal A} + 
\frac{2\,w_\alpha}{m_\alpha^\star}\:{\cal A'}
\ee
and the wave function at the extremity of the lead is 
\be
\psi_{m_\alpha}(0) = 
\frac{2\,w_\alpha}{m_\alpha^\star}\:{\cal A}
+ \frac{2\,w_\alpha^2}{m_\alpha^\star}\:{\cal A'}
\:,\ee
where ${\cal A}=A_1+\cdots+A_{m_\alpha-1}$ and ${\cal A'}=A_{m_\alpha}$.
It is straightforward to see that the matrix involved in the two equations
has an eigenvalue zero associated with the eigenvector 
$(w_\alpha,-1)$. It follows that~:
\be\label{ccwc}
\psi_1(0)=\cdots=\psi_{m_\alpha-1}(0) 
= \frac{1}{w_\alpha}\, \psi_{m_\alpha}(0)
\:.\ee
This equation replaces the continuity condition for the vertices
coupled to the leads.

We now consider the full graph and follow the same lines as in the previous 
section to construct the scattering matrix, by constructing the stationary 
scattering state $\psi^{(\alpha)}(x)$ of energy $E=k^2$ corresponding to a 
plane wave injected from the lead $\alpha$. 
The wave function on the lead connected to 
the vertex $\mu$ is (\ref{wfl}), and (\ref{wfb}) on the internal bonds.
By virtue of (\ref{ccwc}) the continuity condition (\ref{eq1}) is now 
replaced by
\be\label{eqc1}
\delta_{\mu\alpha}  + \Sigma_{\mu\alpha} 
= w_\mu  \: \psi^{(\alpha)}_\mu
\hspace{0.5cm} \mbox{ for } \mu=1,\cdots,L
\:.\ee
The current conservation reads~:
\bea
{\psi_\mu^{(\alpha)}}^*\sum_\beta a_\mb\,\D_{x}\psi^{(\alpha)}_{(\mb)}(\mu)
+ {\psi^{(\alpha)*}_{{\rm lead}\:\mu}}(\mu)\,
\D_x\psi^{(\alpha)}_{{\rm lead}\:\mu}(\mu) &=& 
                             \lambda_\mu \big|\psi_\mu^{(\alpha)}\big|^2
\hspace{0.5cm} \mbox{ for } \mu=1,\cdots,L \\
\sum_\beta a_\mb\,\D_{x}\psi^{(\alpha)}_{(\mb)}(\mu) &=&  
                             \lambda_\mu\psi_\mu^{(\alpha)}
\hspace{0.5cm} \mbox{ for } \mu=L+1,\cdots,V 
\eea
that is now rewritten
\bea
\label{eqc2}
w_\mu
\,\left(\delta_{\mu\alpha}  - \Sigma_{\mu\alpha}\right)
  &=& \sum_\beta M_{\mb} \psi^{(\alpha)}_\beta
\hspace{0.5cm} \mbox{ for } \mu=1,\cdots,L \\ 
\label{eqc3}
0 &=& \sum_\beta M_{\mb} \psi^{(\alpha)}_\beta
\hspace{0.5cm} \mbox{ for } \mu=L+1,\cdots,V
\:.\eea
Equations (\ref{eqc1},\ref{eqc2},\ref{eqc3}) have the same form as 
(\ref{eq3},\ref{eq4}) provided the $L\times V$ matrix $W$ is now defined as~:
\begin{center}
\begin{tabular}{|p{6cm}|}
\hline
\be\label{RES4}
W_\ab = w_\alpha \:\delta_\ab
\:,\ee
\\
\hline
\end{tabular}
\end{center}
with $w_\alpha\in\RR$.
The introduction of tunable couplings between the graph and the leads in 
(\ref{RES2},\ref{RES3}) is thus a simple modification of the matrix $W$ with
(\ref{RES4}).

\subsection*{Resonances}

It is easy to see from the above formalism how the spectrum of resonances
of the graph connected to external leads is related to the eigenvalues 
spectrum of the same isolated graph.
The spectrum of resonances is given by the poles of the scattering matrix,
the real part of the pole being the energy of the resonance and the imaginary
part its width.

\noindent ({\it i}) In the vertex matrix formulation, the poles of $\Sigma$
are the complex zeros of $\det(M+W^{\rm T}W)$.
The matrix $M$ encodes all the information on the isolated graph (topology
of the graph and potential on the bonds) whereas the information on the
way the graph is coupled to the external leads is contained in $W$. If we
turn off the couplings $w_\alpha\to0$, it is clear that we recover the
energies of the isolated graph, solutions\footnote{
Note that in certain cases, the equation $\det M(-k^2)=0$ is not sufficient
to construct all the eigenstates of the graph. However the states missed by 
this equation are found by solving $\det(1-RQ)=0$ for the isolated graph.
Such a situation occurs for example for the complete graph $K_V$ (a graph
with $V$ vertices all connected by bonds of same length) considered 
in \cite{AkkComDesMonTex00}.  
} of $\det M(-k^2)=0$ 
(see \cite{Des00} and remark of section \ref{sec:vm}).

\noindent ({\it ii}) In the arc matrix formulation, the poles are the
zeros of $\det(1-RQ^{\rm int})$. Now the informations on the topology of the
graph and the couplings are mixed in $Q^{\rm int}$ whereas $R$ encodes the
information on the potential. Again, if the couplings are switched off, the
matrix $Q^{\rm int}$ is equal to the matrix $Q$ of the isolated graph
whose spectrum is given by $\det(1-RQ)=0$.

\subsection*{Example 1}

\begin{figure}[!h]
\begin{center}
\includegraphics[scale=0.9]{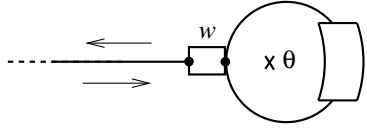}
\end{center}
\caption{Ring with arbitrary coupling $w$ to the lead.}
\label{fig:ring3}
\end{figure}

We compute the scattering matrix of a ring connected to one lead 
(figure \ref{fig:ring3}). This is the situation considered in \cite{But85}.
The result is obtained by replacing (\ref{MQ3}) by (\ref{vce}) in the
calculation we have already done.
A more direct way is to use (\ref{RES2},\ref{RES4}),
$\Sigma=-1 + 2\frac{w^2}{M+w^2}$, the matrix $M$ being given by (\ref{Mring}).

\begin{figure}[!h]
\begin{center}
\includegraphics{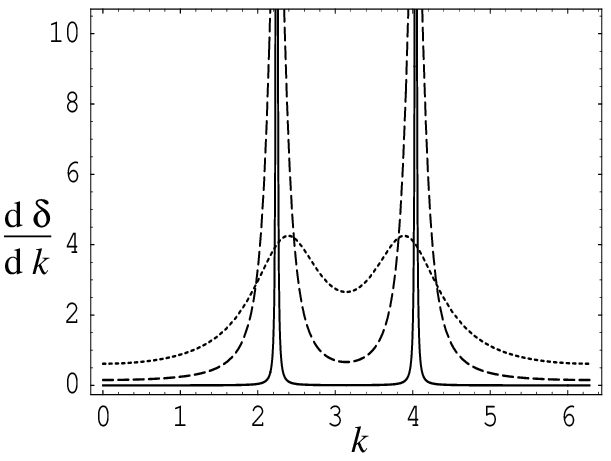}
\hspace{1cm}
\includegraphics{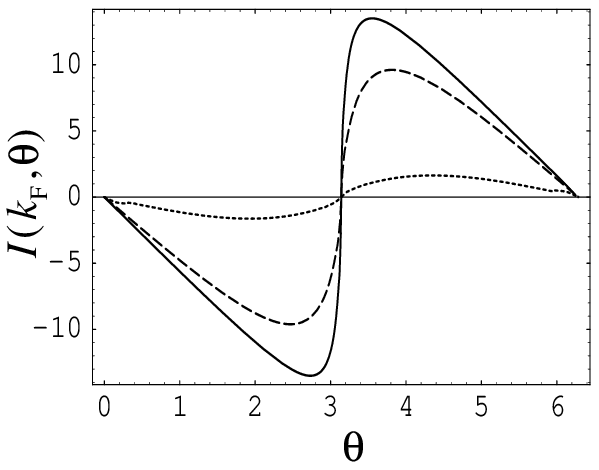}
\end{center}
\caption{{\sc Left}~: $\frac{\D \delta}{\D k}$ for the ring of figure 
\ref{fig:ring3} for different values of the coupling $w$. Dotted lines~: $w=1$,
dashed lines~: $w=0.5$ and full lines~: $w=0.05$. The flux is
$\theta=5\pi/7$ and the length $l=1$.
{\sc Right}~: Persistent current 
$I(k_F,\theta)=\int_0^{k_F}\D k\,2k\,j(k^2,\theta)$
for $w=0.5$ (full line), $w=1$ (dashed lines) and $w=5$ (dotted lines).
We have chosen $k_F l=3\pi$.}
\label{fig:resonances}
\end{figure}

\noindent
If we consider the case of a ring with a potential on the bond like in
figure \ref{fig:ring3}, $M$ is given by
$M=2\I\frac{\cos\Phi-\sqrt{T}\cos\theta}{\sin\Phi-\sqrt{1-T}\cos\varphi}$,
as explained in appendix \ref{sec:loop}.
We obtain $\Sigma = \EXP{\I\delta}$ with~:
\be\label{sringconn}
\cotg\frac{\delta}{2} = w^2\,
\frac{\sin\Phi-\sqrt{1-T}\cos\varphi}
     {2(\sqrt{T}\cos\theta-\cos\Phi)}
\:.\ee
For $w=0$ the ring is disconnected from the arm and the phase shift is constant
($\delta=\pi$).
We now consider the case without a potential on the ring~:
$T=1$, $\Phi=kl$. 
\be\label{sringcb}
\cotg\frac{\delta}{2} = w^2\,
\frac{\sin kl}
     {2(\cos\theta-\cos kl)}
\:.\ee
If $w=1$ we recover the result (\ref{sring2}).
The effect of the parameter $w$ can be seen clearly if we compute
$\frac{\D \delta}{\D k}$ (see figure \ref{fig:resonances})~:
\be
\frac{\D \delta}{\D k} = l\,w^2
\frac{1-\cos\theta\cos kl}
     {(\cos\theta-\cos kl)^2+\frac14 w^4\sin^2kl}
\:.\ee
We now discuss the two ways to decouple the lead from the ring.

\noindent$\bullet$
In the limit $w\to0$, the width of the resonance peaks is 
$\Delta k=\frac{w^2}{2\,l}$, the peaks being centered on the eigen-energies 
of the isolated ring of perimeter $l$~:
$k_n^\pm l=\pm\theta+2n\pi$, with $n\in\NN$ for the sign $+$ and $n\in\NN^*$ 
for the sign $-$. We have~: 
$\frac{\D \delta}{\D k} \simeq2\pi 
 \frac{\Delta k/\pi}{(k-k_n^\pm)^2+\Delta k^2}$ if $k\sim k_n^\pm$.

\noindent$\bullet$
In the limit $w\to\infty$ the three arcs decouple, the ring is open, and  
$\frac{\D \delta}{\D k}$ presents peaks of width 
$\Delta\kappa=\frac{2}{w^2l}(1-(-1)^m\cos\theta)$ centered on the 
eigen-energies of the isolated line of length $l$~:
$\kappa_m l=m\pi$, for $m\in\NN$.

The physical difference of the two limits may also be seen on the persistent
current \cite{AkkAueAvrSha91} (see also \cite{ComMorOuv95})~: 
$j(k^2,\theta)=\frac{1}{2\pi}\drond{}{\theta}\delta(k^2,\theta)$ is the 
current density, {\it i.e.} $j(E,\theta)\,\D E$ is the current of the states
in the energy range $[E,E+\D E[$. We get 
\be
j(k^2,\theta)=
-\frac{1}{2\pi l}\,\frac{\sin\theta\sin kl}{1-\cos\theta\cos kl}
\,\frac{\D \delta}{\D k}
\:.\ee

\noindent$\bullet$
If $w\to0$, the current density presents sharp peaks of alternate signs
at the position of the resonances~: 
$j(k^2,\theta)\simeq \mp\frac{1}{2\pi l}\frac{\D \delta}{\D k}$
for $k\sim k^\pm_n$.
We define the contribution of the peak at $k^\pm_n$ as 
$I_n^\pm=\int_{k^\pm_n-\delta K}^{k^\pm_n+\delta K}\D k\,2k\,j(k^2,\theta)$
with $\delta K$ being a quantity large compare to the resonance width but
small compare the distance between peaks~: 
$w^2\ll \delta K\,l\ll\min{\theta}{\pi-\theta}$.
We immediatly see that 
$I_n^\pm\simeq\frac{4\pi}{l^2}\left(\mp n-\frac{\theta}{2\pi}\right)$~;
we have recovered the persistent current of the isolated ring
${\cal I}_n^\pm=-\drond{}{\theta}(k_n^\pm)^2$.

\noindent$\bullet$
In the limit $w\to\infty$, the current density behaves like~:
$j(k^2,\theta)\propto(k-\kappa_m)\frac{\D\delta}{\D k}$ in the neighbourhood
of the resonance $k\sim \kappa_m$. It follows that the contributions of the
resonance peaks vanish (due to the opening of the ring)~:
$I_m=\int_{\kappa_m-\delta K}^{\kappa_m+\delta K}\D k\,2k\,j(k^2,\theta)
 \simeq0$
(the right part of figure \ref{fig:resonances} indeed shows that the
persistent current decreases as $w$ is increased).

\subsection*{Example 2}

\begin{figure}[!h]
\begin{center}
\includegraphics[scale=0.9]{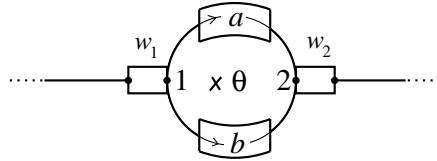}
\end{center}
\caption{Ring pierced by a magnetic flux $\theta$ with potentials on the 
bonds.}
\label{fig:ring2}
\end{figure}

We consider a ring pierced by a flux $\theta$ and connected to two leads 
(see figure \ref{fig:ring2}). This arrangement has been considered in 
several works to study the Aharonov-Bohm oscillations of the conductance of
a normal metal ring~: the authors of \cite{GefImrAzb84} considered a 
particular coupling of the leads whereas \cite{ButImrAzb84} examined 
more general couplings.

The ring is made of two arcs $a$ and $b$.
We use the parameters of appendix \ref{appA} to write the matrix $M$.
Using (\ref{RES3b}) the matrix $M$ is given by~:
\bea
\label{Mr1}
M_{11}         &=& \I\frac{\lambda_1}{k} +
   \I\,\frac{\cos\Phi_a+\sqrt{1-T_a}\sin\varphi_a}
            {\sin\Phi_a-\sqrt{1-T_a}\cos\varphi_a}
 + \I\,\frac{\cos\Phi_b+\sqrt{1-T_b}\sin\varphi_b}
            {\sin\Phi_b-\sqrt{1-T_b}\cos\varphi_b} \\
\label{Mr2}
M_{21}(\theta) &=&
 - \frac{\I\,\sqrt{T_a}\,\EXP{\I\theta/2} }
        {\sin\Phi_a-\sqrt{1-T_a}\cos\varphi_a}
 - \frac{\I\,\sqrt{T_b}\,\EXP{-\I\theta/2}}
        {\sin\Phi_b-\sqrt{1-T_b}\cos\varphi_b} \\ [0.2cm]
\label{Mr3}
M_{12}(\theta) &=& M_{21}(-\theta) \\ [0.2cm]
\label{Mr4}
M_{22}         &=& \I\frac{\lambda_2}{k} +
   \I\,\frac{\cos\Phi_a-\sqrt{1-T_a}\sin\varphi_a}
            {\sin\Phi_a-\sqrt{1-T_a}\cos\varphi_a}
 + \I\,\frac{\cos\Phi_b-\sqrt{1-T_b}\sin\varphi_b}
            {\sin\Phi_b-\sqrt{1-T_b}\cos\varphi_b}
\:.\eea
When several bonds link two vertices $\alpha$ and $\beta$, we have to sum 
the contributions of 
each bond in $M_{\alpha\alpha}$ and $M_{\alpha\beta}$ (see \cite{Pas98} 
and appendix C of \cite{AkkComDesMonTex00}).
Since the two vertices are connected to leads, the matrix $W$ is the 
$2\times2$ diagonal matrix~: 
$W={\rm diag}(w_1,w_2)$. 
We can get the scattering matrix from (\ref{RES2})~:
\be
\Sigma = -1 + \frac{2}{\det(M+W^2)}
\left(
\begin{array}{cc}
w_1^2 M_{22} + w_1^2w_2^2 & -w_1w_2 M_{12} \\
-w_1w_2 M_{21}            & w_2^2 M_{11} + w_1^2w_2^2
\end{array}
\right)
\:.\ee

Now we concentrate ourselves on the case of perfect transmissions through 
the bonds~: $T_{a,b}=1$ and $\Phi_{a,b}=kl_{a,b}$, with $\lambda_{1,2}=0$. 
We have 
\be
\det(M+W^2)= -\frac{2(\cos kl-\cos\theta)}{\sin kl_a\sin kl_b}
+ \I(w_1^2+w_2^2)\frac{\sin kl}{\sin kl_a\sin kl_b} + w_1^2w_2^2
\ee
where $l=l_a+l_b$ is the perimeter of the ring.
If we consider the limit of weak coupling $w_{1,2}\to0$
we can expand the scattering matrix in the neighbourhood of the eigen-energies
of the ring. We obtain the well-known Breit-Wigner form~:
\be
\Sigma_{\ab} \APPROX{k\sim k_n^\pm} -\delta_\ab
+\frac{\I \sqrt{\Delta k_\alpha\Delta k_\beta}\,\EXP{\I\chi_\ab}}
      {k-k_n^\pm + \frac\I2(\Delta k_1+\Delta k_2)}
\:,\ee
where $\Delta k_{1,2}=\frac{w_{1,2}^2}{l}$,
$\chi_{11}=\chi_{22}=0$ and
$\chi_{12}=-\chi_{21}=n\pi\pm \frac12 k_n^\pm(l_a-l_b)$.
Note that a detailed analysis of the resonance structure of the transmission 
probability through the ring has already been done in \cite{ButImrAzb84}.


\section{Summary}

We have given systematic procedures to construct the scattering matrix of 
graphs made of one-dimensional wires on which lives a potential, and 
connected to external leads.

In a first approach we used as basic ingredients a scattering matrix 
(\ref{r1}) describing scattering by the potentials on the bonds 
and a scattering matrix (\ref{q1},\ref{q2},\ref{q3}) providing 
information on the scattering by vertices and coupling to the external leads. 
This approach is quite natural in the sense that we combine the scattering 
matrices of parts of the system to construct the whole scattering matrix
(\ref{RES1}), however it can become cumbersome since we have to deal 
with rather big matrices. 

A way to reduce the problem is to reformulate it in terms of vertex
matrices, which is possible if the scattering at vertices describes
wave functions continuous at the vertices, which allows to deal with vertex
variables instead of arc variables. 

We have described an efficient way to introduce some tunable couplings between 
the leads and the graph (\ref{RES4}), which permits to go continuously 
from a connected graph to an isolate graph.

We have generalized the results known in the absence of potential 
\cite{AvrSad91,KotSmi99,KotSmi00} by adding scattering on bonds and 
allowing to tune the couplings to the external leads.


\section*{Acknowledgements}

One of us (C.T.) would like to acknowledge Marc Bocquet, Alain Comtet, 
Jean Desbois and St\'ephane Ouvry for stimulating discussions.


\begin{appendix}

\mathversion{bold}
\section{Reformulating the matrix $M$ \label{appA}}
\mathversion{normal}

We would like here to use some relations between reflection and transmission
coefficients on a bond to rewrite the result (\ref{RES3}) in terms of
parameters whose physical meanings are more clear. 
In the core of the paper we have considered that the reflection and 
transmission coefficients describe the effect of the scalar potential $V(x)$ 
only. In this appendix we adopt another point of view and consider that 
these coefficients describe the effect of both the scalar potential $V(x)$
and the vector potential $A(x)$.

Due to the unitarity of the
scattering matrix for a given bond $(\ab)$, it follows that the 4 complex 
parameters describing the left ($r_\ab$ and $t_\ab$) and right ($r_\ba$ and 
$t_\ba$) scattering can be parametrized in terms of 4 real parameters~:
\be
\left(
\begin{array}{cc}
r_\ab & t_\ba \\
t_\ab & r_\ba
\end{array}
\right)
=\EXP{\I\Phi_\ab}
\left(
\begin{array}{cc}
\I\EXP{\I\varphi_\ab}\sqrt{1-T_\ab} & \EXP{-\I\chi_\ab}\sqrt{T_\ab} \\ [0.1cm]
\EXP{\I\chi_\ab}\sqrt{T_\ab}        & \I\EXP{-\I\varphi_\ab}\sqrt{1-T_\ab}
\end{array}
\right)
\:.\ee
$\Phi_\ab$ is a global phase. $T_\ab\in[0,1]$ is the transmission probability
through the barrier. In the absence of a magnetic field, we know that the
scattering matrix is symmetric (it is well known that the symmetry of the 
scattering matrix in the presence of a magnetic field ${\cal B}$ is 
$\Sigma(-{\cal B})=\Sigma({\cal B})^{\rm T}$)~; 
it follows 
that we can identify the asymmetric part of the phase of the transmission 
coefficients with the magnetic flux 
\be
\chi_\ab=\theta_\ab
\:.\ee
The last phase $\varphi_\ab$ is related to the asymmetry of the potential
(for $V_{(\ab)}(x)=V_{(\ab)}(l_\ab-x)$ we have $r_\ab=r_\ba$ {\it i.e.}
$\varphi_\ab=0$ or $\pi$). 

Due to these definitions we have the following obvious relations~:
$T_\ab=T_\ba$, $\Phi_\ab=\Phi_\ba$, $\varphi_\ab=-\varphi_\ba$ and we recall
that $\theta_\ab=-\theta_\ba$.

We can now rewrite (\ref{RES3}) in terms of these parameters~:
\begin{center}
\begin{tabular}{|p{13cm}|}
\hline
\bea\label{RES3b}
M_\ab &=&
\I\,\delta_\ab\left(\frac{\lambda_\alpha}{k}
+ \sum_\mu a_{\alpha\mu}
\frac{\cos\Phi_{\alpha\mu}+\sqrt{1-T_{\alpha\mu}}\sin\varphi_{\alpha\mu}}
     {\sin\Phi_{\alpha\mu}-\sqrt{1-T_{\alpha\mu}}\cos\varphi_{\alpha\mu}}
\right)
\nonumber \\ && \hspace{1cm}
- \I\,
a_\ab\frac{\sqrt{T_\ab}\,\EXP{\I\theta_\ab}}
          {\sin\Phi_\ab-\sqrt{1-T_\ab}\cos\varphi_\ab}
\:.\eea
\\
\hline
\end{tabular}
\end{center}
As a by-product, it shows that the matrix $M$ is anti-Hermitian~:
$M^\dagger = -M$.
To end this appendix, we note that if the potential on the bond vanishes 
$V_{(\ab)}(x)=0$, then $T_\ab=1$ and $\Phi_\ab=kl_\ab$.


\mathversion{bold}
\section{Matrix $M$ for a graph with loops\label{sec:loop}}
\mathversion{normal}

\begin{figure}[!h]
\begin{center}
\includegraphics[scale=0.9]{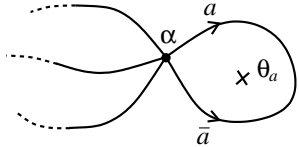}
\end{center}
\caption{A loop at the vertex $\alpha$.}
\label{fig:loop}
\end{figure}

We explain in this appendix how the matrix $M$ is modified when we want to 
describe with the minimum number of vertices a graph possessing loops.
We consider a graph with a loop threatened by a flux $\theta_a$ at the vertex 
$\alpha$ (see figure \ref{fig:loop}). The potential on the arc $a$ of the 
loop is described by four reflection and transmission coefficients~: 
$r_a$, $t_a$ for the arc $a$ and $r_{\bar a}$, $t_{\bar a}$ for the reversed 
arc $\bar a$.

If we follow the lines of section \ref{sec:vm} we can see that only the 
diagonal part of the matrix $M$ (\ref{RES3}) is affected by the loops~:
\be
M_\ab \to M_\ab + \delta_\ab\,M^{\rm loop}_{\alpha\alpha}
\:,\ee
where the contribution of the loop is~:
\bea
M^{\rm loop}_{\alpha\alpha} &=&
 \frac{(1-r_a)(1+r_{\bar a})+t_a t_{\bar a}}
      {(1+r_a)(1+r_{\bar a})-t_a t_{\bar a}}
-\frac{2\,t_a}
      {(1+r_a)(1+r_{\bar a})-t_a t_{\bar a}}
\nonumber \\
&+&
 \frac{(1+r_a)(1-r_{\bar a})+t_a t_{\bar a}}
      {(1+r_a)(1+r_{\bar a})-t_a t_{\bar a}}
-\frac{2\,t_{\bar a}}
      {(1+r_a)(1+r_{\bar a})-t_a t_{\bar a}}
\:.\eea
This result is rather natural~: $M_{\alpha\alpha}$ receives two contributions
from each arc $a$ and $\bar a$ of the kind present in the diagonal elements
of (\ref{RES3}) and since the arc comes back to the same vertex we get 
also two contributions of the kind present in the off-diagonal elements 
of (\ref{RES3}). After simplification we obtain~:
\be
M^{\rm loop}_{\alpha\alpha} = 
2\frac{1 - r_a r_{\bar a} + t_a t_{\bar a} - t_a - t_{\bar a}}
      {(1+r_a)(1+r_{\bar a})-t_a t_{\bar a}}  
\:.\ee

We can also express this contribution with the real parameters introduced 
in the appendix \ref{appA} to describe the scattering by the arc $a$~: 
$\Phi_a=\Phi_{\bar a}$, $T_a=T_{\bar a}$, $\varphi_a=-\varphi_{\bar a}$ and 
$\theta_a=-\theta_{\bar a}$. We obtain~:
\be
M^{\rm loop}_{\alpha\alpha} = 2\I\frac{\cos\Phi_a - \sqrt{T_a}\cos\theta_a}
                           {\sin\Phi_a - \sqrt{1-T_a}\cos\varphi_a}
\:.\ee


\section{Inversion of block matrices}\label{InvMat}

We recall in this appendix a result that can be found in standard textbooks.
Consider the square matrix
\be
{\cal M}=
\left(
\begin{array}{c|c}
A & B \\ \hline
C & D
\end{array}
\right)
\ee
where $A$ and $D$ are square matrices of arbitrary dimensions.
Then~:
\bea
{\cal M}^{-1}
&=&\left(
\begin{array}{c|c}
(A-BD^{-1}C)^{-1}         & -(A-BD^{-1}C)^{-1}BD^{-1} \\ \hline
-D^{-1}C(A-BD^{-1}C)^{-1} & D^{-1}+D^{-1}C(A-BD^{-1}C)^{-1}BD^{-1}
\end{array}
\right) \\
&=&\left(
\begin{array}{c|c}
A^{-1}+A^{-1}B(D-CA^{-1}B)^{-1}CA^{-1} & -A^{-1}B(D-CA^{-1}B)^{-1}\\ \hline 
-(D-CA^{-1}B)^{-1}CA^{-1}              & (D-CA^{-1}B)^{-1}
\end{array}
\right) 
\:.\eea

\end{appendix}



\end{document}